\documentclass[aps,prl,superscriptaddress,showpacs,preprint]{revtex4}
\usepackage{graphicx}
\usepackage{amssymb}
\usepackage{amsmath}

\begin{document}

\bibliographystyle{apsrev}

\title{Strong vortex-antivortex fluctuations in the type II superconducting film}

\author{Yu.N.Ovchinnikov}
\affiliation{Max-Plank Institute for Physics of Complex Systems, Dresden,
D-01187 Germany and Landau Institute for Theoretical Physics, RAS,
Chernogolovka, Moscow District, 142432 Russia}
\author{A. A. Varlamov}
\affiliation{Istituto Nazionale per la Fisica della Materia,
   UdR ``Tor Vergata''/Coherentia,\\
   Via del Politecnico, 1, I-00133 Roma, Italy}

\date{\today}

\begin{abstract}
\medskip
The small size vortex-antivortex pairs proliferation in type II
superconducting film is considered for the wide interval of temperatures
below $T_{c}$. The corresponding contribution to free energy is calculated.
It is shown that these fluctuations give the main contribution to the heat
capacity of the film both at low temperatures and in the vicinity of
transition.
\\
\end{abstract}

\maketitle

The important role of the vortex-antivortex pairs in the 2D phase
transitions is well known. For instance their presence restores the
phenomenon of superconductivity in spite of the destruction of the long
range order by phase fluctuations \cite{BKT}. Let us consider the case of \
dirty superconducting film with the thickness $d\ll \xi _{BCS}\left(
0\right) \sim {\cal D}/T_{c},$ where ${\cal D}=v_{F}l_{tr}/3$ is the
diffusion coefficient and $\tau _{tr}$ is the transport collision time.

Slightly below the BCS critical temperature vortex-antivortex pairing
becomes energetically profitable and the BKT transition to the state with
the finite stiffness takes place. Let us recall that this transition is
determined by the large scale vortex-antivortex pairs, which binding energy
has the logarithmic form. The standard description of the BKT transition is
based on the corresponding logarithmic expression for free energy \cite{BKT}

\begin{equation}
F_{BKT}=\left[ \frac{\pi n_{s2}\left( T\right) }{2m}-2T\right] \ln \frac{R}{%
\xi \left( T\right) },  \label{BKTfreeenergy}
\end{equation}
where $n_{s2}\left( T\right) $ is the superfluid density and $\xi \left(
T\right) $ is the coherence length, which in the vicinity of transition has
the form:

\[
\xi ^{2}\left( T\right) =\frac{\pi {\cal D}}{16T_{c}\tau }
\]
with $\tau =1-T/T_{c}$ as the reduced temperature. Let us underline that
this definition differs twice from the standard GL expression. The
Eq.(\ref{BKTfreeenergy}) is valid for large size vortex-antivortex pairs,
when $R\gg \xi $. Namely these pairs determine the thermodynamical and
transport properties of the $2D$ superconductor in the vicinity of the BKT
transition.

In this letter we will demonstrate that below the immediate vicinity of the
BKT transition, in contrast to the current opinion , the essential role in
fluctuation properties of $2D$ superconductor play the small size
vortex-antivortex pairs. The cornerstone of presented theory consists in the
fact that the energy of the vortex-antivortex pairs turns zero when the
distance between their centers becomes less than $\xi $ \cite{OS99,O01}. As
the consequence such ''cheap'' pairs become ''affordable'' for thermal
fluctuations and their proliferation takes place even at so low temperatures
when other quasiparticles are frozen. In result namely this small size
vortex-antivortex pairs give the dominant contribution with respect to other
fluctuations in the Ginzburg-Landau region below $T_{BKT}$ and even prevail
over the BCS exponential tail in the heat capacity of the $2D$
superconducting system at low temperatures.

In order to take into account the specifics of the mentioned fluctuation
processes let us start from the general expression for the partition function

\[
Z=\int {\frak D}\Delta \left( {\bf r}\right) \int {\frak D}\Delta \left(
{\bf r}\right) ^{\ast }\exp \left\{ -\frac{{F}\left( \Delta \left( {\bf r%
}\right) ,\Delta \left( {\bf r}\right) ^{\ast }\right) }{T}\right\} .
\]
with ${F}\left( \Delta \left( {\bf r}\right) ,\Delta \left( {\bf r}%
\right) ^{\ast }\right) $ as the Ginzburg-Landau functional, which is valid
in the vicinity of $T_{c},$ and then we will generalize the result obtained
for the case of low temperatures. In contrast to the usual Ginzburg-Landau
type long wave-length approximation the calculation of the functional
integral here has to take into account also the vast variety of the order
parameter functions $\Delta _{p}\left( {\bf r}\right) $ corresponding to the
specific realization of the vortex-antivortex pairs in the film. In other
words side by side with the usual Ginzburg-Landau order parameter
fluctuations we will take into account additionally some specific short
wave-length fluctuations.

Let us first separate the partition function $Z_{0}$ of the superconducting
film without fluctuations:
\[
Z=Z_{0}\cdot Z_{\left( fl\right) }.
\]
The partition function $Z_{\left( fl\right) }$ we will calculate in the gas
approximation. Namely we will assume that the main contribution to it
appears from the small pairs neglecting their overlap. Hence

\[
Z_{\left( fl\right) }=Z_{p}^{S/\left[ \pi \xi ^{2}\left( T\right) \right] },
\]
where
\begin{equation}
Z_{p}=\int d\Delta _{\delta }\left( {\bf r}\right) \int d\Delta _{\delta
}\left( {\bf r}\right) ^{\ast }\exp \left\{ -\frac{{F}_{p}\left( \Delta
_{\delta }\left( {\bf r}\right) ,\Delta _{\delta }\left( {\bf r}\right)
^{\ast }\right) }{T}\right\}  \label{Zp}
\end{equation}
is the contribution to the partition function of the isolated single pairs
of all possible sizes $\delta \cdot \xi $, with $0\leq \delta \ll 1$. The
power $S/\left( \pi \xi ^{2}\right) $ in Eq.(\ref{Zp}) takes into account
the combinatorial factor corresponding to the independent formation of such
pairs. \ The order parameter $\Delta _{\delta }\left( {\bf r}\right) $ must
have at the distance $2\delta \xi $ two zeros of the opposite vorticity
(i.e. the total vorticity calculated at large distances is equal to zero).

As the next step we disregard the axial asymmetry of the vortex-anivortex
pair. We will see below that the characteristic size of the pairs which
mainly contribute to the partition function $r_{eff}\ll \xi $ what will
justify our gas approximation $\delta \ll 1.$ The corresponding to Eq.(\ref
{Zp}) free energy functional is
\begin{equation}
{F}_{p}=\nu d\int d^{2}{\bf r}\left[ -\tau |\Delta _{\delta }\left( {\bf %
r}\right) |^{2}+\frac{\pi {\cal D}}{8T_{c}}|\partial _{-}\Delta _{\delta
}\left( {\bf r}\right) |^{2}+\frac{7\zeta \left( 3\right) }{16\pi
^{2}T_{c}^{2}}|\Delta _{\delta }\left( {\bf r}\right) |^{4}\right] .
\label{GLfunc}
\end{equation}
Here $\nu $\ is the density of states and $\partial _{-}=\partial /\partial
{\bf r-}2ie{\bf A}$, $\zeta \left( x\right) $\ is the Riemann zeta-function.
One can demonstrate \cite{OS99} that the energy of $\Delta _{\delta }\left(
{\bf r}\right) $ profile with two zeros at the fixed distance $2\delta \xi $
is majorized by that one of the disc of radius $\delta \xi $ and below we
will estimate the latter.

The GL functional minimization procedure is equivalent to the solution of
the $2D$ Ginzburg-Landau equation with the boundary condition $\Delta
_{\delta }\left( r=\delta \xi \right) =0$.\ Let us assume the order
parameter in the form
\[
\Delta _{\delta }\left( \rho \right) =\Delta _{0}\left( T\right) f\left(
\rho \right)
\]
and write down the dimensionless GL equation:
\begin{equation}
\frac{1}{\rho }\frac{\partial }{\partial \rho }\left( \rho \frac{\partial }{%
\partial \rho }\right) f+\frac{1}{2}f-\frac{1}{2}f^{3}=0.  \label{GLless}
\end{equation}
Here $\rho =r/\xi $ \ is the dimensionless radius, while $\Delta _{0}\left(
T\right) $ is the BCS gap.

Let us investigate the solutions of Eq. (\ref{GLless}) in two limiting
cases: small $\rho \sim \delta \ll 1,$ and large $\rho \rightarrow \infty .$
In the region of small $\rho \sim \delta \ll 1$ the function $f\left( \rho
\right) \rightarrow 0$ and it can be found as the solution of the linearized
Eq. (\ref{GLless}):
\begin{equation}
f\left( \rho \right) =C_{1}J_{0}\left( \rho /\sqrt{2}\right)
+C_{2}N_{0}\left( \rho /\sqrt{2}\right) ,  \label{asssmall}
\end{equation}
where $J_{0}$ and $N_{0}$\ are the Bessel and Neumann functions
correspondingly.

When $\rho \gtrsim 1$, the solution of Eq.(\ref{GLless}) can be looked for
in the form $\ f\left( \rho \right) =1-f_{1}\left( \rho \right) .$
Substituting this expression to Eq.(\ref{GLless}) one finds
\begin{equation}
f\left( \rho \right) =1-f_{1}\left( \rho \right) =1-C_{3}K_{0}\left( \rho
\right) ,  \label{asshigh}
\end{equation}
where $K_{0}$ is the Bessel function. The constant term corresponds to the
usual long wave-length order parameter contribution which will appear in the
result of our consideration side by side with the contributions of the small
vortex-antivortex pairs.

One can see that the asymptotic of Eq.(\ref{asshigh}), being calculated for $%
\rho \gtrsim 1$ nevertheless extends to the region of small values of $\rho
. $ It is why we can match both solutions Eq.(\ref{asssmall}) and Eq.(\ref
{asshigh}) at $\rho \sim \delta \ll 1.$ In order to do this let us use the
expressions for all Bessel functions $J_{0},N_{0}$ and $K_{0}$ at small
arguments \cite{GR}. The comparison of the coefficients at logarithmic term
and of constants gives two equations for coefficients $C_{1,2,3}$ the third
equation is determined by the boundary condition $f\left( \delta \right) =0$
which corresponds to our disc model for the vortex-antivortex pair. Finally

\[
C_{i}=\delta _{i1}+\frac{1}{2\ln \frac{2}{\gamma \delta }}\left\{
\begin{tabular}{l}
$\ln 2$ \\
$\pi $ \\
$2$%
\end{tabular}
\right. .
\]
Here $\gamma =\exp \left( C\right) ,$ where $C=0,577215...$ is the Euler
constant.

Now we can find the energy of disc (vortex-antivortex pair) as the function
of its dimensionless radius $\delta .$ In order to do this one can use the
general expression (\ref{GLfunc}) and the fact that $\Delta =\Delta
_{0}\left( T\right) f$ satisfies the nonlinear Eq. (\ref{GLless}). Using
this and performing the integration of the derivative by parts one can find
the energy of vortex-antivortex pair:
\begin{equation}
{F}_{p}\left( \delta ,T\right) =\frac{B\left( T\right) }{\ln \frac{2}{%
\gamma \delta }}\int_{0}^{\infty }xK_{0}\left( x\right) dx=\frac{B\left(
T\right) }{\ln \frac{2}{\gamma \delta }},  \label{Fp}
\end{equation}
with
\begin{equation}
B\left( T\right) =\frac{\pi ^{2}\nu d{\cal D}\Delta _{0}^{2}\left( T\right)
}{4T_{c}}.  \label{bdef}
\end{equation}
Let us stress that namelly presence of the large logarithm in denominator of
Eq. (\ref{Fp}) does such localized vortex-antivortex type fluctuations of
the order parameter more energetically favorable than the almost homogeneous
(long wave-length) Ginzburg-Landau ones.

Now let us perform the functional integration in Eq.(\ref{Zp}).
Vortex-antivortex pairs proliferate below $T_{c}$ in fluctuation way and in
our model this is equivalent to the account of the discs of the different
radii:
\begin{equation}
Z_{p}=2\int \delta d\delta \exp \left\{ -\frac{{F}_{p}\left( \delta
,T\right) }{T}\right\} .  \label{stat1}
\end{equation}
The integral in Eq. (\ref{stat1}) may be carried out by the method of the
steepest descent what gives
\[
Z_{p}=\frac{4\cdot 2^{1/4}\pi ^{1/2}}{\gamma ^{2}}\left( \frac{B\left(
T\right) }{T}\right) ^{1/4}\exp \left\{ -2\sqrt{\frac{2B\left( T\right) }{T}}%
\right\}
\]
and finally, the corresponding contribution to the free energy is
\begin{equation}
\widetilde{F}_{p}\left( T\right) =-T\ln Z_{p}^{S/\pi \xi ^{2}}=-\frac{TS}{%
\pi \xi ^{2}\left( T\right) }\left[ -2\sqrt{\frac{2B\left( T\right) }{T}}+%
\frac{1}{4}\ln \frac{B\left( T\right) }{T}\right] .  \label{ftilda}
\end{equation}
This expression is valid in the assumption $B\left( T\right) \gg T$ which is
necessary for the applicability of the steepest descent method. Looking on
Eq. (\ref{bdef}) and recalling the definition of the characteristic for
superconducting film $2D$ Ginzburg-Levanyuk number \cite{LV04}
\begin{equation}
Gi_{(2d)}=\frac{21\zeta (3)}{\pi ^{2}}\frac{1}{p_{F}^{2}dl_{tr}}\ll 1
\label{gi}
\end{equation}
one can see that this requirement is equivalent to $\tau \gg Gi_{\left(
2d\right) },$ i.e. our consideration\bigskip\ is valid for temperatures
below the critical region of transition.

At temperatures close to transition, in the GL region \ $1\gg \tau \gg
Gi_{\left( 2d\right) }$ the main temperature dependence of Eq. (\ref{ftilda}%
) originates from functions $\xi \left( T\right) $ and $\Delta _{0}\left(
T\right) .$ Corresponding vortex-antivortex pairs contribution to heat
capacity is
\begin{equation}
C_{p}\left( T\rightarrow T_{c}\right) =-\frac{48ST_{c}}{v_{F}l_{tr}}\frac{%
\partial ^{2}}{\partial \tau ^{2}}\left[ 4\sqrt{\frac{\nu d{\cal D}}{7\zeta
\left( 3\right) }}\tau ^{3/2}-\frac{\tau }{4\pi ^{2}}\ln \frac{2\pi ^{4}\nu d%
{\cal D}\tau }{7\zeta \left( 3\right) }\right] .  \label{chigh}
\end{equation}
One can see that the differentiation of the second term in the square
brackets of the Eq. (\ref{chigh}) gives nothing else as the well known
contribution to heat capacity occurs due to the GL long wave length order
parameter fluctuations \cite{LV04}.
\begin{equation}
\delta C_{\left( fl\right) }^{\left( 2\right) }\left( 1\gg \tau \gg
Gi_{\left( 2d\right) }\right) =\frac{8\pi ^{2}Sd}{7\zeta (3)}\nu T_{c}\left(
\frac{Gi_{(2d)}}{\tau }\right) .  \label{supfl}
\end{equation}
Nevertheless the main fluctuation contribution to the heat capacity of a $2D$
superconducting film ($d\ll \xi \left( {\tau }\right) $) results from the
first term of the Eq. (\ref{chigh}), corresponding to the protagonists of
this letter, i.e. short wave length vortex-antivortex type order parameter
fluctuations:
\begin{equation}
C_{\left( fl\right) }^{\left( v-a\right) }\left( 1\gg \tau \gg Gi_{\left(
2d\right) }\right) =-\frac{48\pi ^{2}Sd}{7\zeta \left( 3\right) }\nu
T_{c}\left( \frac{2Gi_{(2d)}}{\tau }\right) ^{1/2}.  \label{vaheat}
\end{equation}
One can see that the contribution (\ref{vaheat}) dominates over (\ref{supfl}%
) in the entire Ginzburg-Landau region $\tau \gg Gi_{\left( 2d\right) }.$

In the critical region $\tau \lesssim Gi_{\left( 2d\right) }$ the main
contribution to heat capacity belongs to the classical BKT\ large
vortex-antivortex pairs \cite{LV04}:

\[
C^{\left( BKT\right) }\left( \tau \ll Gi_{\left( 2d\right) }\right) \sim
-\nu T_{c}Sd\left( \frac{Gi_{(2d)}}{\tau }\right) ^{3/2}
\]
which matches with Eq. (\ref{vaheat}) at $\tau \sim Gi_{\left( 2d\right) }$
. Let us note that the negative sign of the vortex-antivortex fluctuation
corrections means that the mean field heat capacity jump overestimates its
true value and the vortex-antivortex pairs fluctuation contribution smears
it out.

Now let us turn to discussion of the region of low temperatures.
Unfortunately we cannot write down the exact functional for the
superconductor's free energy here, but it is almost evident that the main
changes in it with respect to the Ginzburg-Landau one at $T\sim T_{c}$\
occur through the different temperature dependence of $\xi \left( T\right) $
and $\Delta _{0}\left( T\right) $ at low temperatures. Hopefully the region
of validity of the Eq. (\ref{ftilda}) in this way can be considerably
extended and, at least as the estimate, we will still use it just\
substituting the GL parameters $\xi \left( T\right) $ and $\Delta _{0}\left(
T\right) $ by their exact BCS values.

The microscopic BCS equation for $\Delta _{0}\left( T\right) $ is well known
\cite{AGD}
\[
\ln \frac{T_{c}}{T}=2\pi T\sum_{\omega _{n}\geq 0}\left( \frac{1}{\omega _{n}%
}-\frac{1}{\sqrt{\omega _{n}^{2}+\Delta _{0}^{2}\left( T\right) }}\right) .
\]
with $\omega _{n}=2\pi T(n+1/2).$ In the limiting cases
\[
\Delta _{0}^{2}\left( T\right) =\left\{
\begin{array}{c}
\frac{8\pi ^{2}T_{c}^{2}}{7\zeta \left( 3\right) }\tau ,\qquad \text{ }\tau
\ll 1 \\
\left( \pi T_{c}/\gamma \right) ^{2},\qquad \text{ }T\ll T_{c}
\end{array}
\right. .
\]
In the region of low temperatures $T\ll T_{c}$ the temperature corrections
to the order parameter are exponentially small by the parameter $\exp \left(
-\Delta _{0}\left( 0\right) /T\right) .$

The coherence length $\xi \left( T\right) $ in the most general case can be
defined as the pole of the linear response operator kernel for the
corresponding correction to modulus of $\Delta $ \cite{ALO69}. In the
simplest case of dirty superconductor the corresponding equation takes form
\begin{equation}
\sum_{\omega _{n}>0}\left[ \frac{\omega _{n}^{2}}{\left[ \omega
_{n}^{2}+\Delta _{0}^{2}\left( T\right) \right] \left( \sqrt{\omega
_{n}^{2}+\Delta _{0}^{2}\left( T\right) }-{\cal D}/\left[ 2\xi ^{2}\left(
T\right) \right] \right) }-\frac{1}{\sqrt{\omega _{n}^{2}+\Delta
_{0}^{2}\left( T\right) }}\right] =0.  \nonumber
\end{equation}
In the limiting cases of low and high temperatures
\[
\xi ^{2}\left( T\right) =\left\{
\begin{array}{c}
\frac{\pi {\cal D}}{16T_{c}\tau },\text{ \qquad }T\rightarrow T_{c} \\
\frac{{\cal D}}{2\theta \Delta _{0}\left( 0\right) },\text{\qquad }T\ll T_{c}
\end{array}
\right. ,
\]
where $\theta =0.76595$ is the solution of the equation

\[
\frac{\pi }{4}=\sqrt{1-\theta ^{2}}\left[ \frac{\pi }{2}-\arctan \left(
\sqrt{\frac{1-\theta }{1+\theta }}\right) \right] .
\]

Now we are ready to calculate the contribution of the vortex-antivortex
fluctuations to the heat capacity $C^{\left( v-a\right) }\left( T\ll
T_{c}\right) $of the$2D$ superconducting film ($d\ll \xi \left( {0}\right) $%
) at low temperatures. The values $\xi \left( T\right) $ and $\Delta
_{0}\left( T\right) $ here can be assumed to be temperature independent: $%
\xi \left( T\right) \approx \xi \left( 0\right) =\frac{{\cal D}}{2\theta
\Delta _{0}\left( 0\right) }$ and $\Delta _{0}\left( T\right) \approx \Delta
_{0}\left( 0\right) =\pi T_{c}/\gamma $ and differentiation of the Eq. (\ref
{ftilda}) results in
\begin{equation}
\frac{C^{\left( v-a\right) }\left( T\ll T_{c}\right) }{\Delta C\left(
T_{c}\right) }=\frac{7\pi \zeta \left( 3\right) \theta }{8(l_{tr}p_{F})}%
\frac{1}{\gamma ^{2}}\sqrt{\frac{T_{c}}{T}\frac{3l_{tr}}{d}}=0.87\left(
\frac{Gi_{(2d)}}{t}\right) ^{1/2},  \label{clow}
\end{equation}
where $t=T/T_{c}\ll 1$ and
\[
\Delta C=\frac{8\pi ^{2}Sd}{7\zeta \left( 3\right) }\nu T_{c}
\]
is the BCS jump of heat capacity at $T_{c}$.

Contribution (\ref{clow}) requires a special discussion. First of all let us
stress its power dependence on temperature and unusual growth with the
decrease of temperature. Comparison of \ Eq. (\ref{clow}) with the BCS
expression for heat capacity of superconductor shows that already at $%
T\lesssim T_{c}/\ln \left[ Gi_{(2d)}^{-1}\right] $ the former exceeds the
latter, i.e. nonmonotoneous temperature dependence of a heat capacity can be
expected. Naturally the formal divergence of Eq. (\ref{clow}) at very low
temperatures has not be taken seriously: at low temperatures the
quasiclassical approximation used for evaluation of the vortex-antivortex
energy \cite{OS99,O01} breaks down and these excitations probably acquire
some small gap $\left( \sim \Delta ^{2}/E_{F}\right) $ in the spectrum (in
complete analogy with the excitations in the vortex core in $2D$
superconductors \cite{CDM64,LO98} ).

In conclusion let us summarize the results obtained. We have demonstrated
that the possibility of small vortex-antivotrex pairs proliferation in $2D$
superconducting film results in appearance of the specific contribution to
its heat capacity. Being gapless such excitations result in the appearance
of the non-exponential heat capacity of superconducting film at low
temperatures which turns to dominate the BCS term. At high temperatures the
small vortex-antivortex pairs contribution to heat capacity dominates over
all other fluctuation corrections in the entire GL region below $T_{c}.$
Finally, only within the critical region the large BKT type
vortex-antivortex pairs become of the first importance and determine the
thermodynamics of the $2D$ superconducting film.

The authors are grateful to A. Larkin for valuable discussion. Yu.N.O.
acknowledges the financial support of the CARIPLO foundation (Italy) CRDF
under the proposal 13124 (USA) and the Russian Foundation for Basic
Research. A.V. acknowledges the financial support of the FIRB ''Coherence
and transport in nanostructures'' and COFIN 2003.

\end{document}